\documentclass[10pt,letterpaper]{article}

\usepackage{cogsci}
\usepackage{pslatex}
\usepackage{apacite}
\usepackage{amsmath}
 \usepackage{graphicx}

\title{Online network topology shapes personal narratives and hashtag generation}

 \author{\textbf{J. Hunter Priniski} \\ Department of Psychology\\ UCLA 
 \And
 \textbf{Bryce Linford} \\ Department of Psychology\\ UCLA 
 \And
\textbf{Sai Krishna} \\ Information Sciences Insitute\\ USC
\AND \textbf{Fred Morstatter} \\ Information Sciences Institute\\ USC 
\And \textbf{Jeff Brantingham} \\ Department of Anthropology\\ UCLA 
\And \textbf{Hongjing Lu} \\ Department of Psychology\\ UCLA
}

\begin{document}

\maketitle

\begin{abstract}
While narratives have shaped cognition and cultures for centuries, digital media and online social networks have introduced new narrative phenomena. With increased narrative agency, networked groups of individuals can directly contribute and steer narratives that center our collective discussions of politics, science, and morality. We report the results of an online network experiment on narrative and hashtag generation, in which networked groups of participants interpreted a text-based narrative of a disaster event, and were incentivized to produce matching hashtags with their network neighbors. We found that network structure not only influences the emergence of dominant beliefs through coordination with network neighbors, but also impacts participants' use of causal language in their personal narratives. 


\end{abstract}

\section{Introduction}
The internet has remapped how people interpret and discuss events. Digital technology enables organizing efforts that transcend geographic limitations, leading to some of the largest demonstrations in history \cite{dawson_hashtag_2020,yang_narrative_2016}. However, these same platforms can create information environments that foster extremism, hate, and anti-democratic ideals. Echochambers, where online communities consume media that confirms their beliefs and identities, are breeding grounds for unreliable information and conspiracy theories \cite{sasahara_social_2021}. Empirical research is necessary to understand how networked environments shape belief formation at both individual and group levels, so as to better control the dynamics of information spread, and to possibly mitigate the harm of misinformation and social segregation.

Empirically investigating networked behavior is a challenging task because the narratives arising in online contexts are sprawling and unwieldy \cite{tangherlini_automated_2020}, largely due to the dynamic and complex nature of modern social media environments. For instance, the interactivity of digital media and online social networks allows for a new sense of \textit{narrative agency} \cite{yang_narrative_2016}. Through the production and sharing of social media data, users can embed personal narratives and express their point of view on an event or social issue \cite{boyd_tweet_2010}. From these low-level interactions, a collective narrative emerges within a group of individuals, who bring their own prior beliefs and goals when characterizing an event or issue, which directly shapes the narrative shared by a network \cite{dawson_hashtag_2020}.

Hashtags are a potent force for narrative interaction on social media; they allow users to tag personal narratives and contribute to online discourse by indexing their produced content with proxy topic labels. Previous research has shown that hashtags are concise representations of the narratives \cite{giaxoglou2018jesuischarlie, dawson_hashtag_2020}, and connect spatially disorganized groups according to the content of their narratives and goals \cite{papacharissi_affective_2015, papacharissi_affective_2016,howard_democracys_2013}. Across an online network, hashtags categorize social media discourse for effective indexing and search, and can allow interpersonal signaling and sense making between instances \cite{papacharissi_affective_2015, papacharissi_affective_2016}.  

Previous research on hashtags has primarily focused on how they are used in real-world (i.e., ``scale-free'') networks, by focusing on the linguistic and semantic structure of popular hashtags \cite{booten_hashtag_2016}, or modeling the dynamics underlying their spread online \cite{cunha_analyzing_2011, lin_bigbirds_2013}. For instance, hashtags fall into one of two categories. \textit{Focal hashtags} tag posts with broad semantic topics to relate posts to broader discussions and movements across an onine network. A second set of \textit{individualistic hashtags} make the distribution of hashtags heavy-tailed, as they co-occur with focal hashtags while allowing users use to signal personal narratives \cite{booten_hashtag_2016}. Furthermore, hashtags generally fall into a ``winner' or ``loser'' category \cite{lin_bigbirds_2013}, in that while many hashtags initially compete for popularity, only a small set of hashtags will persist to allow for broader narrative collaboration across the network. It is unclear how endorsed narratives shape the production of hashtags and how hashtag dynamics are sensitive to network structures. To address these important questions, empirical experiments must mirror the interaction structure and media of online environments to effectively account for how groups engage with narratives in real-world networks, and how group behaviors are affected by an individual's representation of the underlying event.

One approach to studying narrative interaction is to run social network experiments, where a group of participants are placed in a social network and interact with one another. This paradigm allows network structure to be manipulated under experimental control. Social network experiments have historically used relatively simple materials to measure how varying social network structure (e.g., node connectivity) influences the adoption of normative behaviors. For example, Centola and Baronchelli \citeyear{centola_spontaneous_2015} asked participants to coordinate on naming an image of a face with network neighbors. They found that interactions within \textit{homogeneously-mixed networks}—fully-connected networks where each participant is linked to every other participant—support the emergence of normative behaviors (i.e., the full network aligning on a single name), while \textit{spatially-embedded networks}, where each participant is linked only to a handful of neighbors, did not. Here, we extend this well-established network experiment design by using naturalistic narrative materials and interaction behaviors akin to those in real-world online networks. We hypothesize that repeated interactions in localized neighborhoods will allow groups to coordinate more effectively, but neighborhoods spanning a fully-connected network will be more likely to produce dominant behaviors. To test these hypotheses, we developed fine-grained measures of behavioral coherence at both the level of local coordination between pairs of participants, and at the level of global convergence across the full network. In addition, we applied an NLP model to compare people's causal language use in personal narratives before and after networked interaction. 

\section{Online Network Experiment on Personal Narrative and Hashtag Generation}

\subsection{Participants}
We sampled a total of $N = 420$ participants from the Prolific and SONA subject pools, and placed them into one of ten network conditions. Conditions vary the size of a network ($N = 20, 50, 100$) as well as its connectivity (homogeneously-mixed/fully-connected; spatially-embedded/ring-like). We collected six network runs for $N=20$ (three runs per network structure), and single runs for each structure of $N=50$ and $N=100$. Participants $N = 20$ and $N = 50$ conditions were sampled using Prolific. For the $N = 100$ condition, we recruited undergraduates in the Department of Psychology at UCLA through SONA . We posted initial recruitment surveys a week prior to each run in SONA and a few hours prior to each  run in Prolific. Participants who received the most points at the end of the experiment received an additional $\$10$ bonus.

\subsection{Materials}
\begin{figure}
    \centering
    \includegraphics[width=\linewidth, trim={2.5cm 5cm 2.5cm 5cm}, clip]{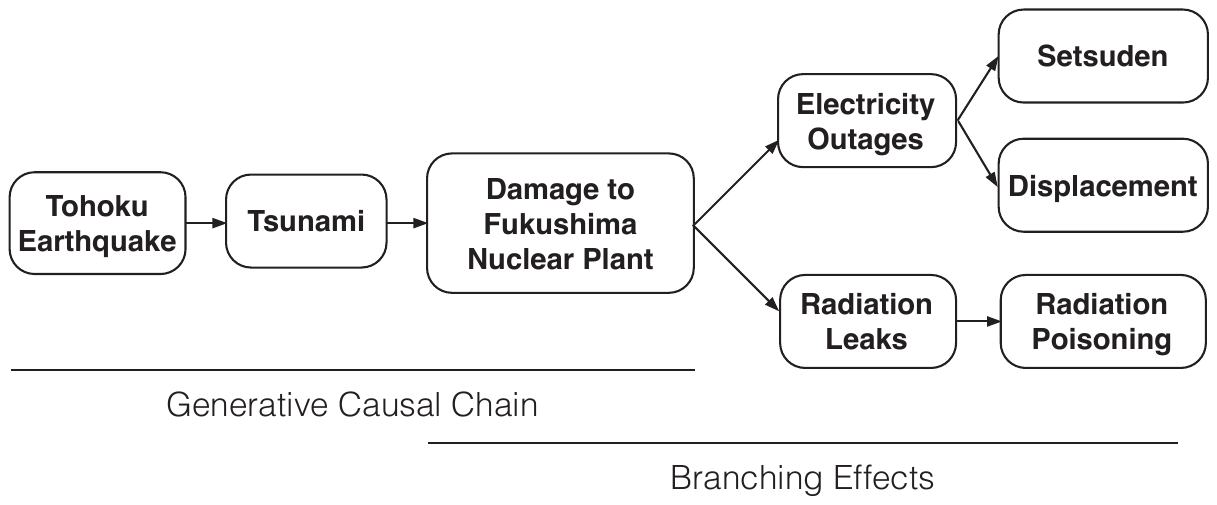}
    \caption{\textbf{Causal model communicated in the nuclear disaster narrative}. This diagram is just for illustration purposes, participants did not see this diagram. They read a four-paragraph narrative describing how the T$\overline{\mathrm{o}}$hoku earthquake triggered a massive tidal wave that damaged the Fukushima Nuclear Power Plant, resulting in electricity outages, radiation leaks and poisoning, human displacement, and \textit{Setsuden}, a national energy-saving holiday.}
    \label{fig:Fuku-cause-model}
\end{figure}

Across all network conditions, participants first read a four-paragraph narrative description of the Fukushima nuclear disaster. The narrative explains how a large earthquake triggered a tsunami that caused damage to a nuclear reactor and resulted in radiation leaks, population displacement, and an energy-saving movement ``Setsuden''. We selected this narrative based on a pilot study demonstrating that it resulted in the most diverse set of hashtags within a set of tested narratives related to natural and financial disasters. This is likely because the narrative describes a rich set of causal relations (a generative causal chain producing a branching common cause sequence) and included both negative (e.g., displacement, poisoning) and positive effects (e.g., energy saving movement). Fig. \ref{fig:Fuku-cause-model} illustrates the causal structure of the Fukushima disaster narrative. 

\subsection{Experimental Design and Procedure}

\subsubsection{Network Experiment Method}
We used the open-source framework OTree written in Python \cite{chen_otreeopen-source_2016}, and hosted experiments on a Linux server. Participants joined the experiment through a Qualtrics survey that directed participants to the network experiment. 

Our social network experiment proceeded in three steps. First, we randomly assigned each participant as a player in a network that defined who may interact with whom on a given trial. Second, we assigned interactions between individual participants on each trial. Third, we rewarded participants based on the outcome of their interactions. We can specify this process using graph theory notation. The first step is to initialize a fixed graph $G(N, E)$, defined by a set $N$ nodes representing individual participants connected through an edge set $E$. We discuss below the specific graph structures used. The second step iterates over $T$ trials. On a given trial $t \in T$, connection (edge) configurations follow mixing participants randomly within a participant's neighborhood. The third step is to identify and reward coordinated behavior. If the response from participant $ n_i $ on trial $ t $ is $r^t_{i}$, then  participants \( n_i \) and \( n_j \) coordinate if $r^t_{i} = r^t_{j}$.

\begin{figure}
    \centering
    \includegraphics[width=1\linewidth]{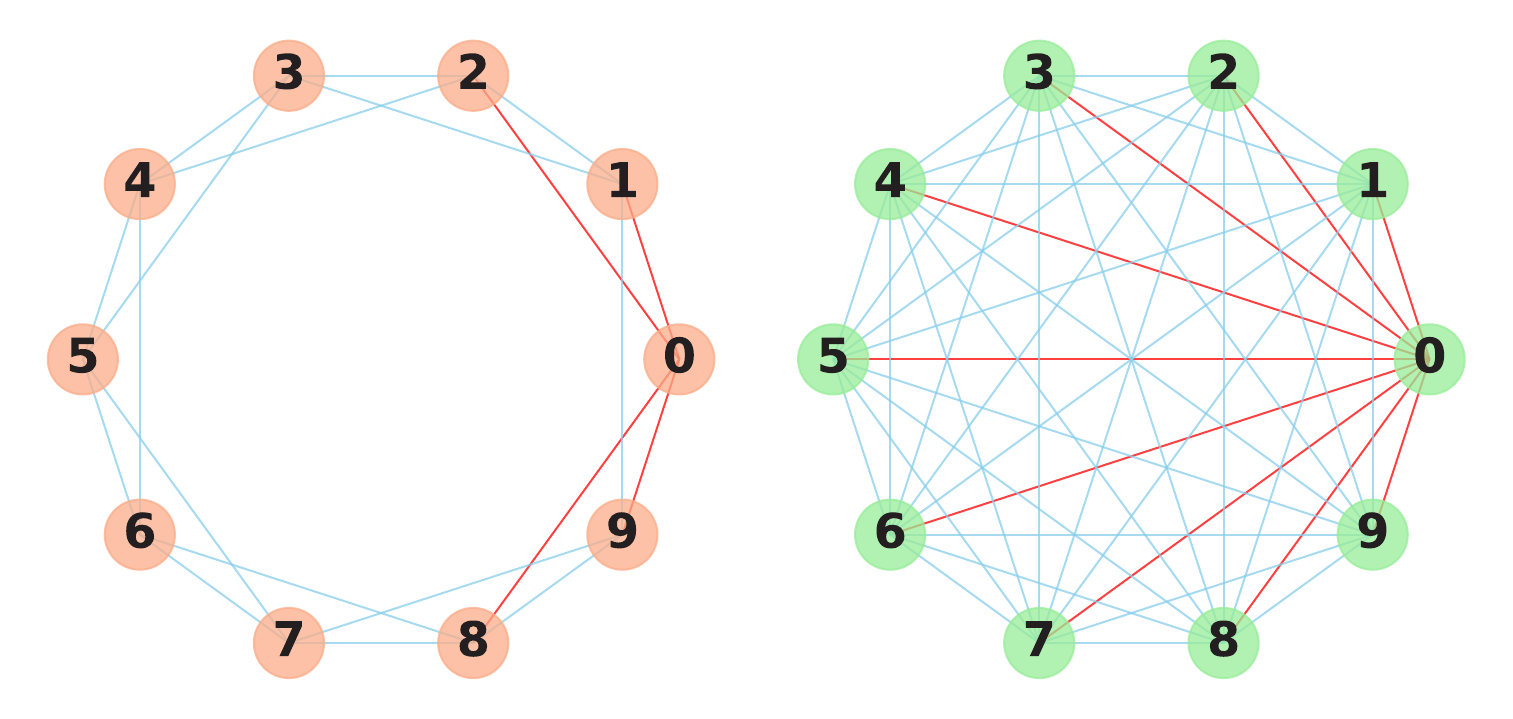}
    \caption{\textbf{Two neighborhood structures tested in this experiment:} Spatially-embedded (\textbf{left}) and homogeneously-mixed (\textbf{right}) networks with $N = 10$ nodes. Red edges represent the neighborhoods for a hypothetical node 0 in both networks. As a network's size grows, the diameter of spatial networks grow whereas homogeneous networks maintain a diameter of $1$.}
    \label{fig:net-structures}
\end{figure}

\subsubsection{Procedure}
The experiment consisted of three blocks: a pre-interaction block, a networked interaction block (described above), and a post-interaction block. This three-block design allowed us to assess behavioral dynamics during the networked interaction block, in addition to examining whether networked interaction can shift beliefs from pre- to post-interaction blocks.  

In the \textbf{pre-interaction block}, participants read a four paragraph narrative describing the Fukushima nuclear disaster, and then were asked to write a ``tweet'' (within a 140 word limit) and ten hashtags characterizing the events described in the narrative. 

In the \textbf{networked interaction block}, participants joined a network experiment through real-time interaction on the online platform. Participants were assigned to one of six experimental conditions based on the size of the network ($N = 20$; $50$; $100$) and network structure (spatially-embedded and homogeneously-mixed; see Fig. \ref{fig:net-structures}). Regardless of network size, nodes in spatial networks have a consistent neighborhood size $k = 4$, meaning each participant would interact with four other participants in the entire experiment. Neighborhood size in homogeneous networks is $N - 1$, as each participant can interact with any of the remaining participants. A consequence is that the network diameter (i.e., the largest geodesic distance in the connected network) was consistently $1$ in all tested homogeneous networks, but grows as a function of size in spatial networks. Both of these features of network topology (i.e., size and diameter) uniquely influence  the emergence of shared behavior in online networks \cite{anagnostopoulos2012online}.

The networked interaction block consisted of 40 trials, where participants interacted with their partners based on the edge structure in the assigned network. On each trial, participants were instructed to write a single hashtag describing the narrative they read in the pre-interaction block. After participants submitted their hashtag response, they were then presented with a new page showing their own hashtag response, their partner's hashtag response, whether they received a point for matching responses with their partner, and their cumulative reward point. 

Following networked interactions, participants entered a \textbf{post-interaction block} where they wrote one more ``tweet" for the same narrative and another ten hashtags describing the event. Before completing the experiment, they provided demographic information. 


\subsection{Results and Discussion}

\subsubsection{Global connections support emergence of dominant responses}

We fit a Bayesian generalized linear model (GLM) to predict how the two network structures (spatially-embedded vs homogeneously-mixed structure) support the emergence of a dominant hashtag response. We assume that the proportion of participants who produced the dominant hashtag on trial $t$ follows a Beta distribution, a commonly used distribution to predict proportion values \cite{mcelreath_statistical_2016}; we used uninformative priors (i.e., $\mathcal{N}(0,10)$) over regression coefficients. Specifically, the GLM model predicted the proportion value as a function of trial number (i.e., $Trial$) interacted with network structure ($Spatial$ vs $Homogeneous$), while controlling for network $size$.  Here, we simply predict the proportion of players in a network producing the dominant response on a given trial, which could be different hashtags across different trials for a single run. 

\begin{figure}
    \centering
    \includegraphics[width=1\linewidth]{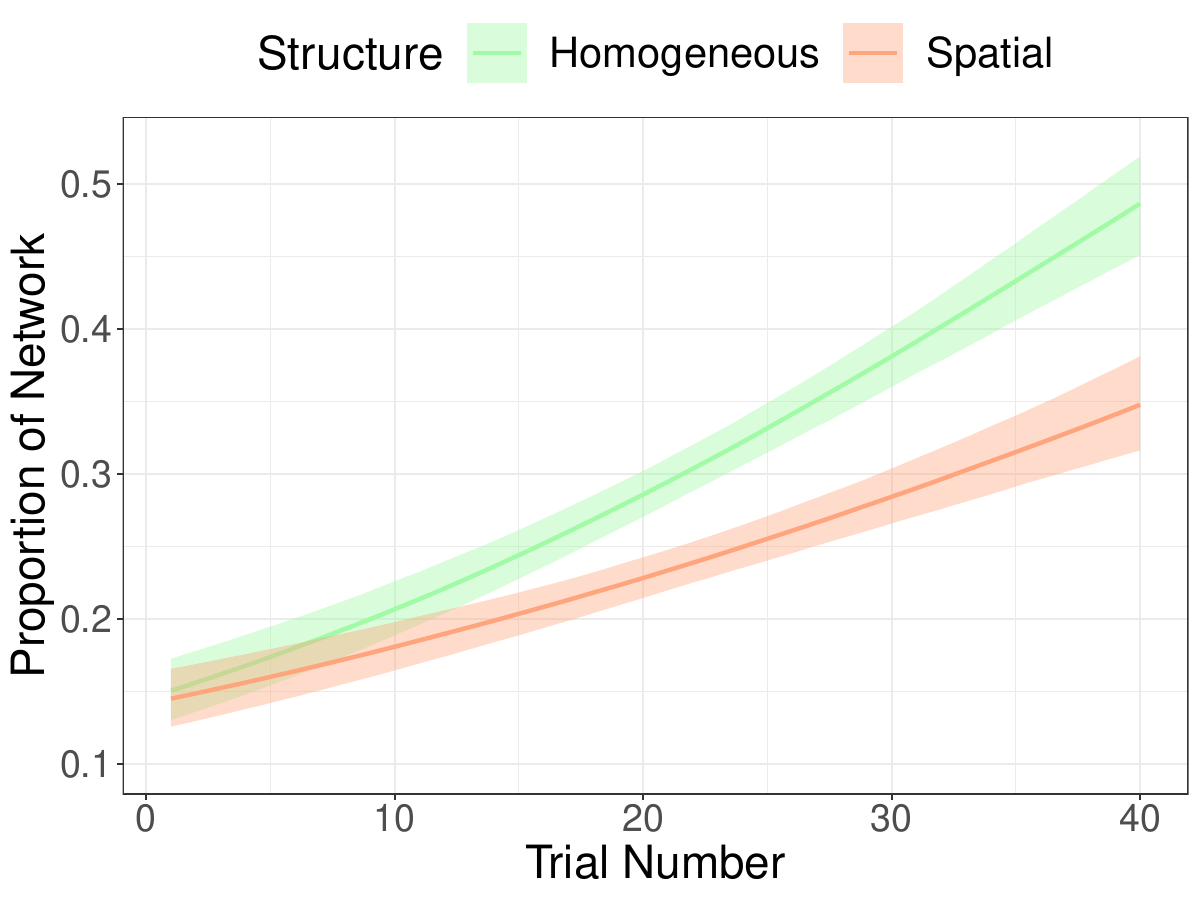}
    \caption{\textbf{Proportion of participants generating a dominant hashtag as a function of trials across network structures.} Curves represent marginal effects of the linear model, and error bars represent $95\%$ Bayesian credible intervals. Homogeneously-mixed networks show faster emergence of dominant responses due to information aggregation across network connections.}
    \label{fig:dom_fig}
\end{figure}

As shown in Fig. \ref{fig:dom_fig}, shared responses emerge from networked interactions in both network structures  ($\beta_{Trial} = 0.04$, $95\%$ CI $[0.04,0.05]$), however dominant responses emerge more easily in homogeneously-mixed networks than spatially-embedded networks ($\beta_{Trial \times Spatial} = -0.01$, $95\%$ CI $[-0.02,-0.00]$). We found that within the confines of a given experimental run (i.e., $40$ interaction trials), the shared responses emerge more quickly in smaller network sizes than larger ones  ($\beta_{Size} = -0.01$, $95\%$ CI $[-0.01,-0.00]$) (note that this effect represents the additive effect of increasing network size by one). These results suggest that  network structure and size are important characteristics to determine the adoption of shared beliefs, and replicates previous findings from behavioral economics and the computational social sciences \cite{golub2010naive, centola_spontaneous_2015}. 



\subsubsection{Network topology affects entropy dynamics of response distribution}

Responses from all players in a network produce a distribution of generated hashtags on each trial. The variability of hashtags over time captures the degree of coherence of individual responses across all participants in a network. Hence, entropy is a concise measure of response variability across a network \cite{avolio_comprehensive_2019,hallett_codyn_2016}; the lower the entropy, the more similar the set of responses from all players. We computed the change in entropy of the response distributions across each network run, and fit a Gaussian generalized linear model to predict a network's entropy as a function of time and structure (and their interaction), while controlling for a network's size. 

\begin{figure}[ht]
    \centering
    \includegraphics[width=1\linewidth]{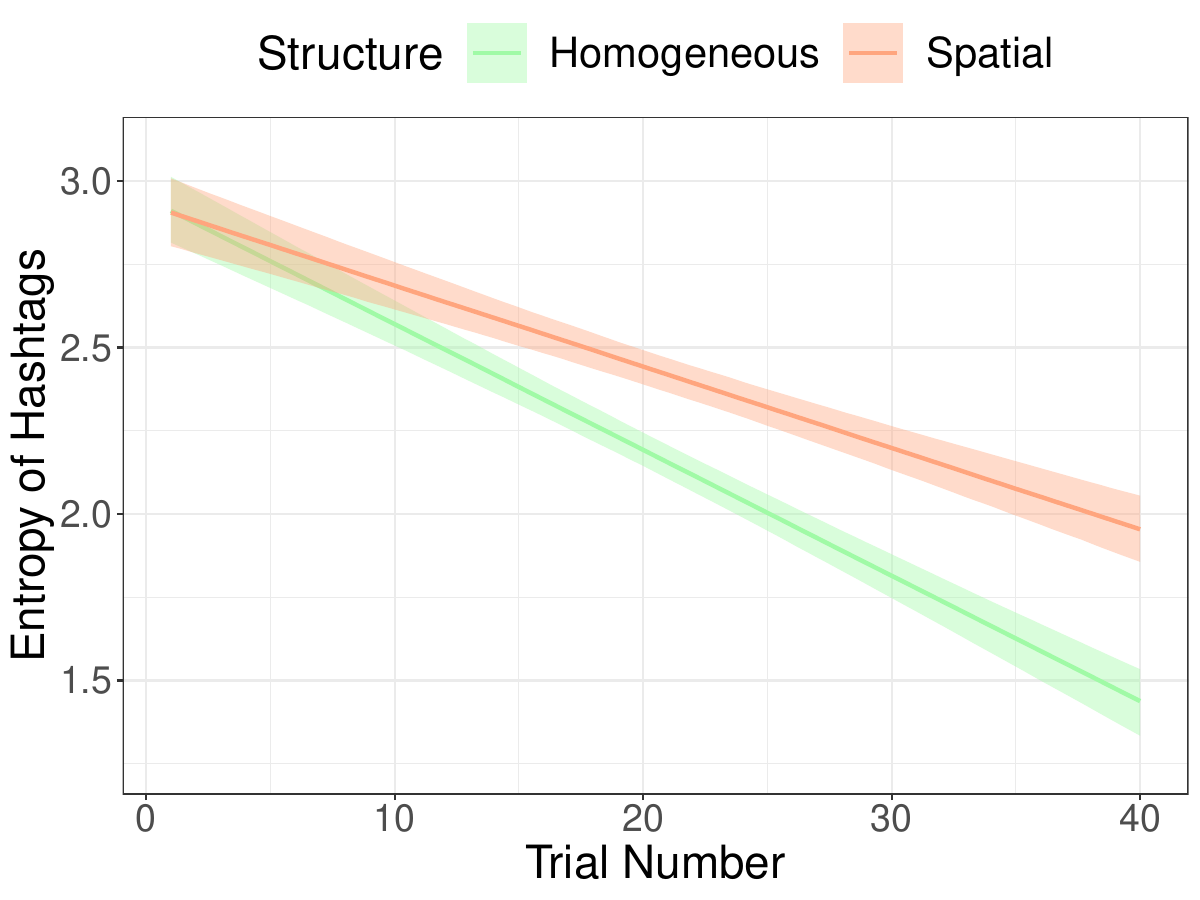}
    \caption{\textbf{Entropy of hashtag responses decrease over time across network structures.} Lower values imply greater coherence of responses across participants because generated hashtags are more similar. However, entropy of responses decrease more rapidly in homogeneously-mixed networks due wider communication of information.}
    \label{fig:entropy_fig}
\end{figure}

As shown in Fig. \ref{fig:entropy_fig}, the GLM model shows that the starting entropy is similar across network structures ($\beta_0 = 2.59$, $95\%$ CI $[2.47, 2.71]$). Furthermore, the entropy of hashtag responses decreases  over time ($\beta_{Trial} = 0.04$, $95\%$ CI $[-0.04, -0.03]$), however entropy decreases at a faster rate in homogeneously-mixed networks than spatially-embedded networks ($\beta_{Spatial} = 0.01$, $95\%$ CI $[0.01, 0.02]$). Because larger networks have more participants contributing responses, we also found a positive effect of network size on entropy ($\beta_{Size} = 0.01$, $95\%$ CI $[0.01, 0.01]$). While a single dominant response may struggle to emerge in spatially-embedded networks, responses are still cohering within local neighborhoods, which in turn decreases the entropy (less variability) of hashtag responses, but at a slower rate than homogeneously-mixed networks. Because separate local-neighborhood groups can align on different hashtags in spatially-embedded networks, this finding is consistent with echochambers found in online networks. We discuss this idea in more detail at the end of this section (see Fig. \ref{fig:colormap}). 


\subsubsection{Local connections support coordination within neighborhoods}

\begin{figure}[ht]
    \centering
    \includegraphics[width=\linewidth]{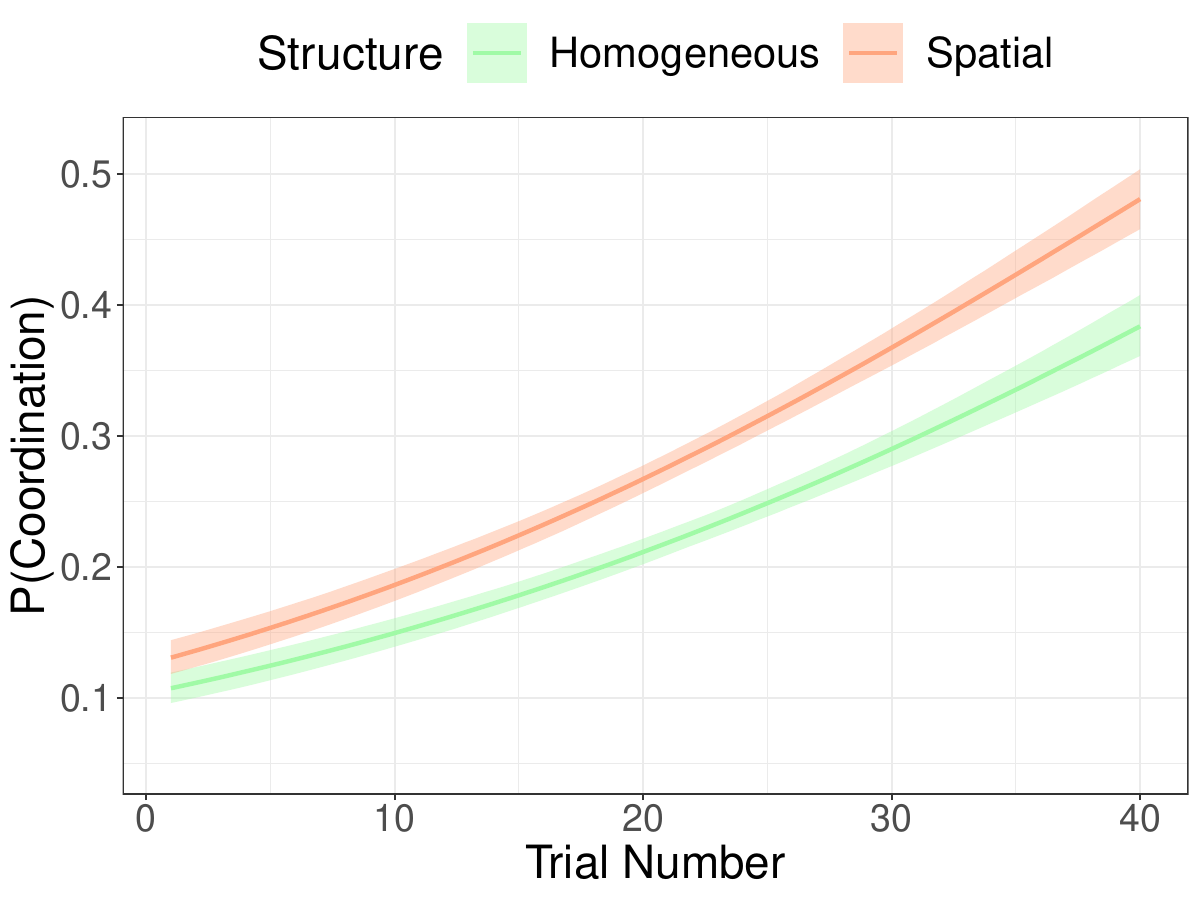}
    \caption{\textbf{Emergence of coordinated behavior across network structures.}} 
    \label{fig:coord_analysis}
\end{figure}

Connections across a network support the emergence of a dominant belief. However, increasing the connectivity of a node decreases the amount of times that any given pair of players can coordinate. In the spatially-embedded, partner players coordinate ten times across forty trials, regardless of total network size; whereas in homogeneously-mixed networks repeated interactions significantly decrease as a function of network size (i.e., number of players in the network). To assess how coordination dynamics varied across network structures, we fit a Bernoulli generalized linear model to predict if a pair of participants coordinated on trial $t$ interacted with network structure over time (controlling for the size of the network). 

As shown in Fig. \ref{fig:coord_analysis}, participants in both neighborhood structures learned to coordinate with their networked neighbors as trials progressed ($\beta_{Trial} = .04$, 95\% CI [0.00, 0.04]), however those in spatially-embedded networks coordinated more effectively than they did in homogeneously-mixed networks ($\beta_{Spatial} = .02$, 95\% CI [0.01, 0.03]).

\begin{figure}[ht]
    \centering
    \includegraphics[width=1\linewidth]{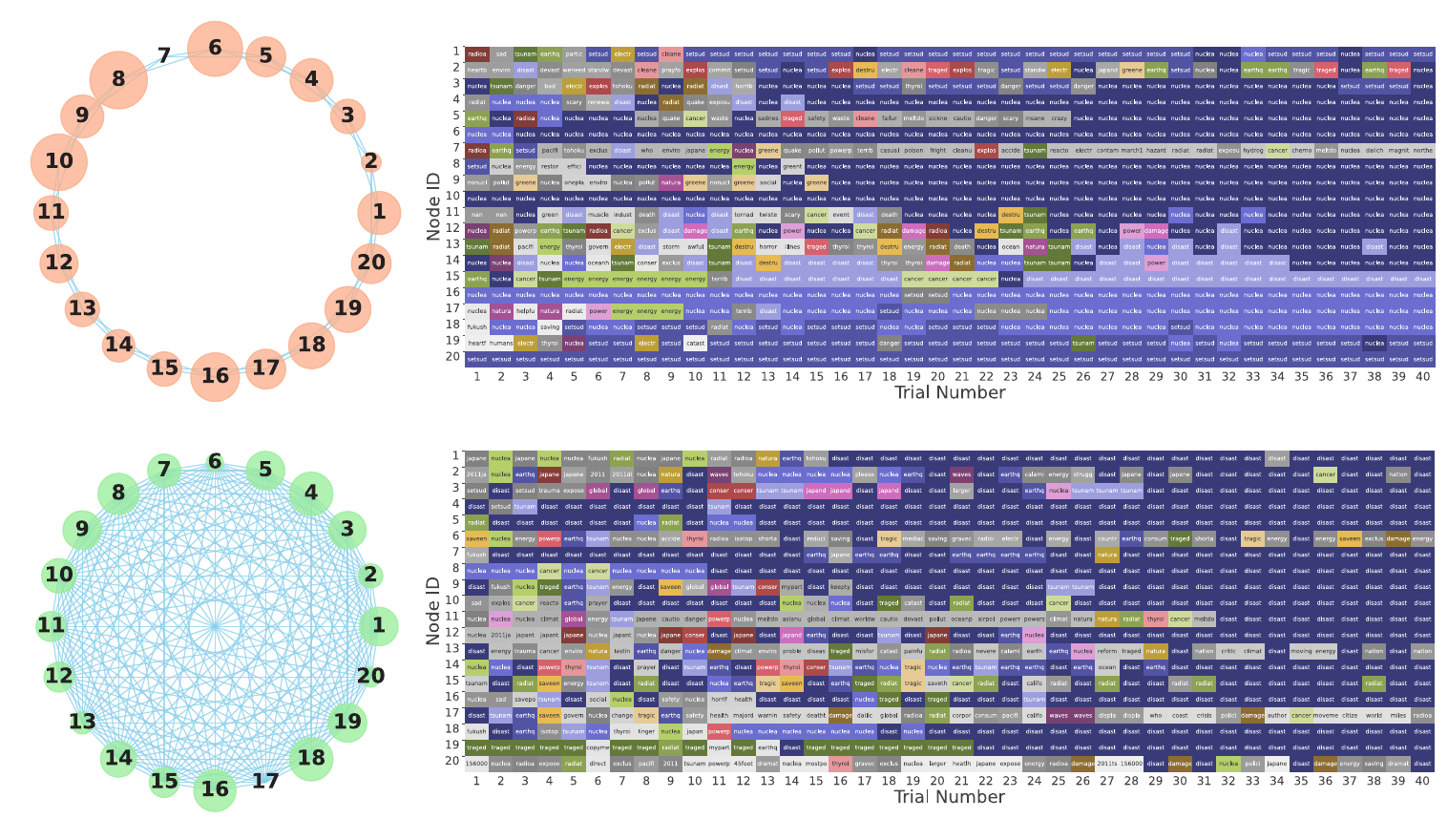}
    \caption{\textbf{Rewards and colormaps of hashtag responses across a single $N = 20$ run.} Top panel shows results for a spatially-embedded network, the bottom panel is from a homogeneously-mixed network. \textbf{Left:} Network structure with player nodes sized by participants' final rewards for coordinating (range (top) 0-25; (bottom) 0-19). \textbf{Right:}
    Colormap of individual responses, rows represent individual participants' set of responses, columns represent trials. Cells encode the first five letters of the generated hashtag.}
    \label{fig:colormap}
\end{figure}

Local coordination can result in separable neighborhoods coordinating on different hashtags. Fig. \ref{fig:colormap} depicts the full array of responses for a single run of both $N=20$ network structures. Different local groups align on different hashtags in the spatially-embedded network. For  example, 16-18 adopted \#NuclearDisaster, while 2-6 and 8-14  aligned on \#Nuclear; and nodes 19, 20, and 1 aligned on \#Setsuden. The emergence of separate, localized groups coordinating on different hashtags likely hinders a dominant response from emerging in these networks, and limits the decrease of entropy as shown in the earlier analysis of global network coherence. The other $N=20$ runs and network sizes $N=50,100$ display similar results. Participants in localized clusters received high rewards for coordinating within their partners in the local clusters (as indicated by the node size in the networks to the left of the color maps). Although participants were not coordinating as effectively in homogeneously-mixed networks, network topology still supported the emergence of a dominant response globally. 

This result suggests that a latent form of information aggregation leads to the emergence of dominant hashtag, rather than  being directly due to participants coordinating effectively in the local neighborhood \cite{golub2010naive}. Furthermore, note how in Fig. \ref{fig:colormap}, participants $7$ (top) and $17$ (bottom) both received zero reward for coordinating, and continually generated new hashtags over course of the experiment. Participants are rewarded by adopting shared behaviors that encourage a shift towards consensus, and those that don't learn to exploit environmental regularities are not rewarded. 

\subsubsection{Probing causal representation change following networked interactions}

Before and after networked interaction, participants wrote ``tweets'' describing the narrative. Natural language processing (NLP) methods applied to these documents can illuminate which parts of the narrative people focused on when describing the events in a brief format such as tweets. Because causal relations are central to narrative representation \cite{zwaan_dimensions_1995, morrow_updating_1989}, we analyzed the causal claims that participants made in the tweet documents. 

Priniski et al. \citeyear{priniski_pipeline_2023} developed a Large Language Model (LLM) system that identifies and extracts causal claims expressed in text documents. The model identifies spans of words serving as inputs to explicitly stated causal relations. Both the cause and effect events, and the underlying causal relation (i.e., a causal trigger) must be explicitly stated for the algorithm to identify the causal claim. The extracted claims are then co-referenced based on BERT embeddings of the identified entities \cite{devlin_bert_2019}, to produce clusters of semantically similar topics, termed ``causal topics''. The model additionally encodes the direction of the stated causal relationships linking any two clusters.  

\subsubsection{Networked interaction shifts causal language expressed in personal narratives}
To assess if participants expressed more causal language following networked interaction, we fit a hurdle Poisson model to predict the number of causal claims a participant produced at a given phase of interaction. The hurdle Poisson model consists of a logistic classification step for identifying tweets with no identified causal claims, and then a Poisson distribution estimates the counts for remaining documents. Because some participants may be more prone to generate causal reports than others, we fit the model with a random intercept for subject.  Effects are to be interpreted as cumulative log odds which describes the expected number of claims in each interaction phase and network condition. 

The hurdle parameter predicts that around $54$\% of documents contained zero causal claims ($hu = .54$, 95\% CI [0.51, 0.57]). The intercept equals the expected log count of causal claims before networked interaction in a homogeneously-connected network ($\beta_0 = .19$, 95\% CI [-0.04, 0.39]), which equates to a mean of $1.19$ claims. Interestingly, after networked interaction, the expected count increases to around $1.61$, which is more than pre-interaction counts ($\beta_{Post} = .30$, 95\% CI [0.08, 0.52]). Participants in spatially-embedded networks produced slightly fewer claims than those in homogeneously-mixed networks, however there was not a credible difference ($\beta_{Spatial} = -.12$, 95\% CI [-0.34, 0.09]). 

\begin{figure}
    \centering
    \includegraphics[width=\linewidth]{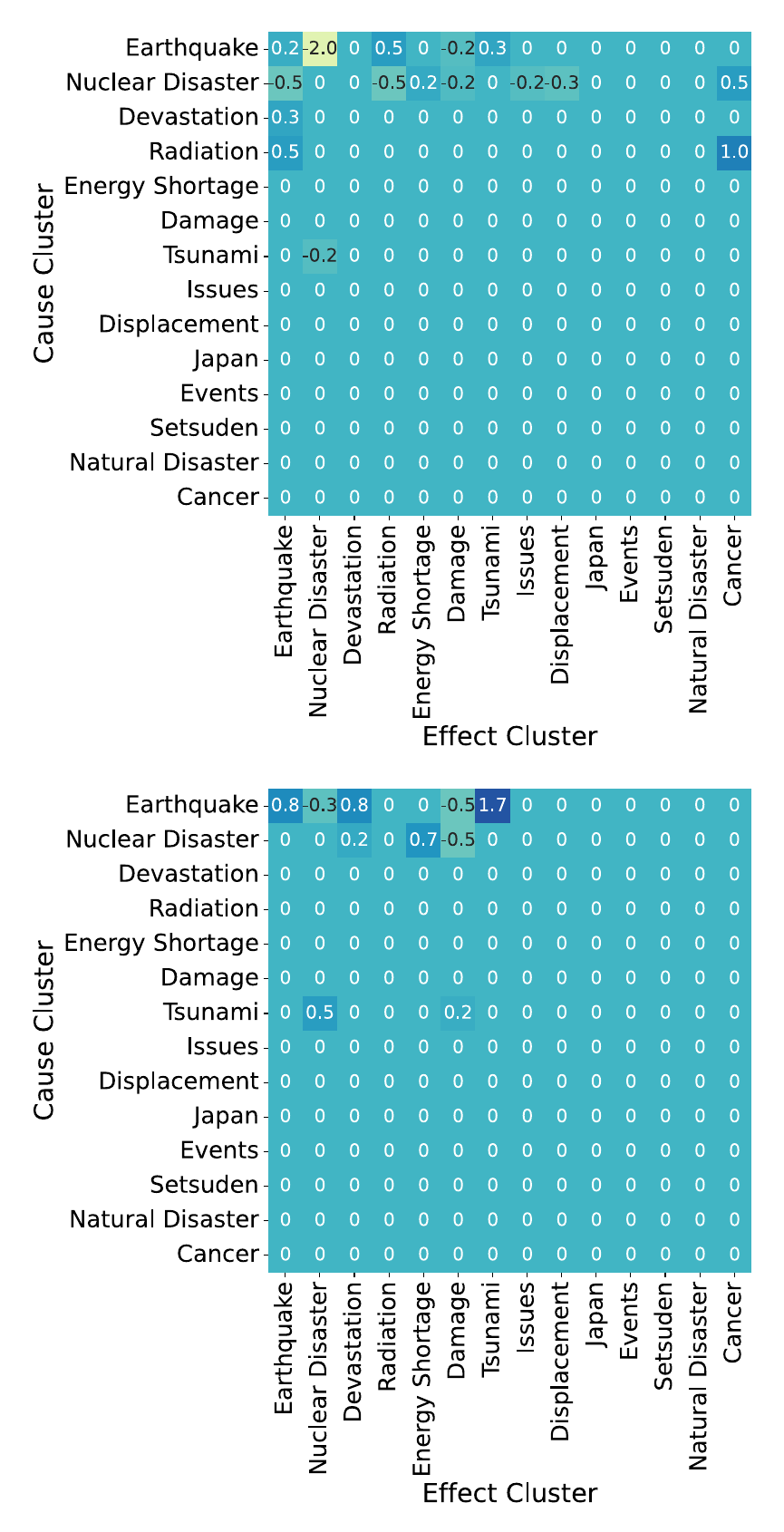}
    \caption{\textbf{Shift in causal language following networked interaction.} \textbf{Top:} spatially-embedded networks; \textbf{bottom:} homogeneously-mixed networks.  Cells represent average difference scores of claims instantiating each causal topic across network structures. Cell $i,j$ represents documents claiming that topic $i$ caused $j$. Positive values indicate more documents expressed that causal relationship after interaction, negative values indicate less causal claims in the tweet documents after network interaction.}
    \label{fig:heatmap}
\end{figure}

Finer-grained analyses of the content of participants' causal claims can reveal what causal topics are most salient in the narrative, and how network structure may shift an individual's representation of the narrative's causal content. We compared differences in claims made after and before interaction to highlight causal relations that may have been enhanced by interaction. As shown in Fig. \ref{fig:heatmap}, the model identified fourteen causal clusters, plus a non-topic category of claims that couldn't be clustered based on their embeddings. All causal events expressed in the narrative appear as clusters in the corpus (for reference see Fig. \ref{fig:Fuku-cause-model}), reflecting a tendency to use causal claims when summarizing events, even in short-formatted messages such as tweets. Because causal relations are directional (causes produce effects), Fig. \ref{fig:heatmap} shows the direction of the relationships linking any two topics. Rows denote clusters used as a cause and columns denote clusters used as an effect. Cell values are the averaged differences across networked conditions in the number of claims expressing that causal relation after interaction relative to before (i.e, post - pre-interaction). Due to the global aggregation of information in homogeneously-mixed networks, participants in these networks generate tweet messages including a smaller set of causal relations centering around the initial causal chain in the narrative. For example, for the groups with homogeneously-mixed interactions, the causal link \textit{earthquake $\rightarrow$ tsunami} had increased by $1.7$ documents after interaction, and the subsequent causal link \textit{tsunami $\rightarrow$ nuclear disaster} had increased by $.5$. However, the enhancement of this causal chain is much weaker for the groups with spatially-embedded interactions; the difference scores are $.3$ and $-.2$, respectively.

\section{Discussion}
We examined how the similarity between a full set of hashtags across a network, and the similarity between pairs of coordinating nodes is sensitive to a network's neighborhood topology. We replicated the finding that global connections facilitate the emergence of dominant behaviors, while repeated interactions within localized neighborhoods promotes higher rewards based on coordination, with separable communities producing different hashtag responses. 

These findings have direct implications for understanding the onset of echochambers and belief polarization in online social networks. High reward values in the spatial networks proxy social reinforcement for stating one's beliefs online. People will increase their confidence in their stated beliefs when those in their neighborhood reaffirm their responses, despite different beliefs reported from others in different communities. After separate clusters of people begin coordinating on different responses, it will become more difficult to find common ground once beliefs are solidified with one's groups \cite{kahan_polarizing_2012}. Our results suggest that one way to increase global agreement within a online network is to structure neighborhoods as to encourage interactions across a wider-array of nodes in the network. Even if participants don't directly coordinate their beliefs with their direct contacts, access to this information could have an aggregate impact on the global consensus of the network. 

In addition to analyzing hashtag reporting during networked interaction, we measured shifts in causal language use before and after networked interaction. We found that more causal topics emerged after network interactions, especially more in homogeneous structure than in spatial structure. These results suggest that network interactions have the potential to deepen people's understanding of causality in complex events.  Future NLP analyses of ``tweets'' should parse a wider-array of semantic relations to model shifts in \textit{situation models}, the memory representations people build when processing text-based narratives \cite{morrow_updating_1989, zwaan_construction_1995, zwaan_situation_1998}. This effort could elucidate how networked interactions impact people's narrative representations, and illuminate mechanisms for encouraging healthier discourse online. 

\subsection{Acknowledgements}
This work was funded in part by the AFOSR MURI grant No. FA9550-22-1-0380 and the DARPA Army Research Office (ARO), under Contract No. W911NF-21-C-0002. We thank Yiling Yun, Yiting Wang, and Chole Ji for their help in running the experiments. 


\bibliographystyle{apacite}

\setlength{\bibleftmargin}{.125in}
\setlength{\bibindent}{-\bibleftmargin}

\bibliography{CogSci_Template}

\begin{thebibliography}{}

\bibitem [\protect \citeauthoryear {%
Anagnostopoulos%
, Becchetti%
, Castillo%
, Gionis%
\BCBL {}\ \BBA {} Leonardi%
}{%
Anagnostopoulos%
\ \protect \BOthers {.}}{%
{\protect \APACyear {2012}}%
}]{%
anagnostopoulos2012online}
\APACinsertmetastar {%
anagnostopoulos2012online}%
\begin{APACrefauthors}%
Anagnostopoulos, A.%
, Becchetti, L.%
, Castillo, C.%
, Gionis, A.%
\BCBL {}\ \BBA {} Leonardi, S.%
\end{APACrefauthors}%
\unskip\
\newblock
\APACrefYearMonthDay{2012}{}{}.
\newblock
{\BBOQ}\APACrefatitle {Online team formation in social networks} {Online team formation in social networks}.{\BBCQ}
\newblock
\BIn{} \APACrefbtitle {Proceedings of the 21st international conference on World Wide Web} {Proceedings of the 21st international conference on world wide web}\ (\BPGS\ 839--848).
\PrintBackRefs{\CurrentBib}

\bibitem [\protect \citeauthoryear {%
Avolio%
\ \protect \BOthers {.}}{%
Avolio%
\ \protect \BOthers {.}}{%
{\protect \APACyear {2019}}%
}]{%
avolio_comprehensive_2019}
\APACinsertmetastar {%
avolio_comprehensive_2019}%
\begin{APACrefauthors}%
Avolio, M\BPBI L.%
, Carroll, I\BPBI T.%
, Collins, S\BPBI L.%
, Houseman, G\BPBI R.%
, Hallett, L\BPBI M.%
, Isbell, F.%
\BDBL {}Wilcox, K\BPBI R.%
\end{APACrefauthors}%
\unskip\
\newblock
\APACrefYearMonthDay{2019}{}{}.
\newblock
{\BBOQ}\APACrefatitle {A comprehensive approach to analyzing community dynamics using rank abundance curves} {A comprehensive approach to analyzing community dynamics using rank abundance curves}.{\BBCQ}
\newblock
\APACjournalVolNumPages{Ecosphere}{10}{10}{e02881}.
\PrintBackRefs{\CurrentBib}

\bibitem [\protect \citeauthoryear {%
Booten%
}{%
Booten%
}{%
{\protect \APACyear {2016}}%
}]{%
booten_hashtag_2016}
\APACinsertmetastar {%
booten_hashtag_2016}%
\begin{APACrefauthors}%
Booten, K.%
\end{APACrefauthors}%
\unskip\
\newblock
\APACrefYearMonthDay{2016}{}{}.
\newblock
{\BBOQ}\APACrefatitle {Hashtag Drift: Tracing the Evolving Uses of Political Hashtags Over Time} {Hashtag drift: Tracing the evolving uses of political hashtags over time}.{\BBCQ}
\newblock
\BIn{} \APACrefbtitle {Proceedings of the 2016 {CHI} Conference on Human Factors in Computing Systems} {Proceedings of the 2016 {CHI} conference on human factors in computing systems}\ (\BPGS\ 2401--2405).
\newblock
\APACaddressPublisher{}{{ACM}}.
\PrintBackRefs{\CurrentBib}

\bibitem [\protect \citeauthoryear {%
Boyd%
, Golder%
\BCBL {}\ \BBA {} Lotan%
}{%
Boyd%
\ \protect \BOthers {.}}{%
{\protect \APACyear {2010}}%
}]{%
boyd_tweet_2010}
\APACinsertmetastar {%
boyd_tweet_2010}%
\begin{APACrefauthors}%
Boyd, D.%
, Golder, S.%
\BCBL {}\ \BBA {} Lotan, G.%
\end{APACrefauthors}%
\unskip\
\newblock
\APACrefYearMonthDay{2010}{{\APACmonth{01}}}{}.
\newblock
{\BBOQ}\APACrefatitle {Tweet, {Tweet}, {Retweet}: {Conversational} {Aspects} of {Retweeting} on {Twitter}} {Tweet, {Tweet}, {Retweet}: {Conversational} {Aspects} of {Retweeting} on {Twitter}}.{\BBCQ}
\newblock
\BIn{} \APACrefbtitle {2010 43rd {Hawaii} {International} {Conference} on {System} {Sciences}} {2010 43rd {Hawaii} {International} {Conference} on {System} {Sciences}}\ (\BPGS\ 1--10).
\PrintBackRefs{\CurrentBib}

\bibitem [\protect \citeauthoryear {%
Centola%
\ \BBA {} Baronchelli%
}{%
Centola%
\ \BBA {} Baronchelli%
}{%
{\protect \APACyear {2015}}%
}]{%
centola_spontaneous_2015}
\APACinsertmetastar {%
centola_spontaneous_2015}%
\begin{APACrefauthors}%
Centola, D.%
\BCBT {}\ \BBA {} Baronchelli, A.%
\end{APACrefauthors}%
\unskip\
\newblock
\APACrefYearMonthDay{2015}{}{}.
\newblock
{\BBOQ}\APACrefatitle {The spontaneous emergence of conventions: {An} experimental study of cultural evolution} {The spontaneous emergence of conventions: {An} experimental study of cultural evolution}.{\BBCQ}
\newblock
\APACjournalVolNumPages{Proceedings of the National Academy of Sciences}{112}{7}{1989--1994}.
\PrintBackRefs{\CurrentBib}

\bibitem [\protect \citeauthoryear {%
Chen%
, Schonger%
\BCBL {}\ \BBA {} Wickens%
}{%
Chen%
\ \protect \BOthers {.}}{%
{\protect \APACyear {2016}}%
}]{%
chen_otreeopen-source_2016}
\APACinsertmetastar {%
chen_otreeopen-source_2016}%
\begin{APACrefauthors}%
Chen, D\BPBI L.%
, Schonger, M.%
\BCBL {}\ \BBA {} Wickens, C.%
\end{APACrefauthors}%
\unskip\
\newblock
\APACrefYearMonthDay{2016}{{\APACmonth{03}}}{}.
\newblock
{\BBOQ}\APACrefatitle {{oTree}—{An} open-source platform for laboratory, online, and field experiments} {{oTree}—{An} open-source platform for laboratory, online, and field experiments}.{\BBCQ}
\newblock
\APACjournalVolNumPages{Journal of Behavioral and Experimental Finance}{9}{}{88--97}.
\PrintBackRefs{\CurrentBib}

\bibitem [\protect \citeauthoryear {%
Cunha%
\ \protect \BOthers {.}}{%
Cunha%
\ \protect \BOthers {.}}{%
{\protect \APACyear {2011}}%
}]{%
cunha_analyzing_2011}
\APACinsertmetastar {%
cunha_analyzing_2011}%
\begin{APACrefauthors}%
Cunha, E.%
, Magno, G.%
, Comarela, G.%
, Almeida, V.%
, Gonçalves, M\BPBI A.%
\BCBL {}\ \BBA {} Benevenuto, F.%
\end{APACrefauthors}%
\unskip\
\newblock
\APACrefYearMonthDay{2011}{{\APACmonth{06}}}{}.
\newblock
{\BBOQ}\APACrefatitle {Analyzing the {Dynamic} {Evolution} of {Hashtags} on {Twitter}: a {Language}-{Based} {Approach}} {Analyzing the {Dynamic} {Evolution} of {Hashtags} on {Twitter}: a {Language}-{Based} {Approach}}.{\BBCQ}
\newblock
\BIn{} M.~Nagarajan\ \BBA {} M.~Gamon\ (\BEDS), \APACrefbtitle {Proceedings of the {Workshop} on {Language} in {Social} {Media} ({LSM} 2011)} {Proceedings of the {Workshop} on {Language} in {Social} {Media} ({LSM} 2011)}\ (\BPGS\ 58--65).
\newblock
\APACaddressPublisher{Portland, Oregon}{Association for Computational Linguistics}.
\PrintBackRefs{\CurrentBib}

\bibitem [\protect \citeauthoryear {%
Dawson%
}{%
Dawson%
}{%
{\protect \APACyear {2020}}%
}]{%
dawson_hashtag_2020}
\APACinsertmetastar {%
dawson_hashtag_2020}%
\begin{APACrefauthors}%
Dawson, P.%
\end{APACrefauthors}%
\unskip\
\newblock
\APACrefYearMonthDay{2020}{}{}.
\newblock
{\BBOQ}\APACrefatitle {Hashtag narrative: Emergent storytelling and affective publics in the digital age} {Hashtag narrative: Emergent storytelling and affective publics in the digital age}.{\BBCQ}
\newblock
\APACjournalVolNumPages{International Journal of Cultural Studies}{23}{6}{968--983}.
\PrintBackRefs{\CurrentBib}

\bibitem [\protect \citeauthoryear {%
Devlin%
, Chang%
, Lee%
\BCBL {}\ \BBA {} Toutanova%
}{%
Devlin%
\ \protect \BOthers {.}}{%
{\protect \APACyear {2019}}%
}]{%
devlin_bert_2019}
\APACinsertmetastar {%
devlin_bert_2019}%
\begin{APACrefauthors}%
Devlin, J.%
, Chang, M\BHBI W.%
, Lee, K.%
\BCBL {}\ \BBA {} Toutanova, K.%
\end{APACrefauthors}%
\unskip\
\newblock
\APACrefYearMonthDay{2019}{{\APACmonth{05}}}{}.
\newblock
\APACrefbtitle {{BERT}: {Pre}-training of {Deep} {Bidirectional} {Transformers} for {Language} {Understanding}.} {{BERT}: {Pre}-training of {Deep} {Bidirectional} {Transformers} for {Language} {Understanding}.}
\newblock
\APACaddressPublisher{}{arXiv}.
\newblock
\APACrefnote{arXiv:1810.04805 [cs]}
\PrintBackRefs{\CurrentBib}

\bibitem [\protect \citeauthoryear {%
Giaxoglou%
}{%
Giaxoglou%
}{%
{\protect \APACyear {2018}}%
}]{%
giaxoglou2018jesuischarlie}
\APACinsertmetastar {%
giaxoglou2018jesuischarlie}%
\begin{APACrefauthors}%
Giaxoglou, K.%
\end{APACrefauthors}%
\unskip\
\newblock
\APACrefYearMonthDay{2018}{}{}.
\newblock
{\BBOQ}\APACrefatitle {\#JeSuisCharlie? Hashtags as narrative resources in contexts of ecstatic sharing} {\#jesuischarlie? hashtags as narrative resources in contexts of ecstatic sharing}.{\BBCQ}
\newblock
\APACjournalVolNumPages{Discourse, context \& media}{22}{}{13--20}.
\PrintBackRefs{\CurrentBib}

\bibitem [\protect \citeauthoryear {%
Golub%
\ \BBA {} Jackson%
}{%
Golub%
\ \BBA {} Jackson%
}{%
{\protect \APACyear {2010}}%
}]{%
golub2010naive}
\APACinsertmetastar {%
golub2010naive}%
\begin{APACrefauthors}%
Golub, B.%
\BCBT {}\ \BBA {} Jackson, M\BPBI O.%
\end{APACrefauthors}%
\unskip\
\newblock
\APACrefYearMonthDay{2010}{}{}.
\newblock
{\BBOQ}\APACrefatitle {Naive learning in social networks and the wisdom of crowds} {Naive learning in social networks and the wisdom of crowds}.{\BBCQ}
\newblock
\APACjournalVolNumPages{American Economic Journal: Microeconomics}{2}{1}{112--149}.
\PrintBackRefs{\CurrentBib}

\bibitem [\protect \citeauthoryear {%
Hallett%
\ \protect \BOthers {.}}{%
Hallett%
\ \protect \BOthers {.}}{%
{\protect \APACyear {2016}}%
}]{%
hallett_codyn_2016}
\APACinsertmetastar {%
hallett_codyn_2016}%
\begin{APACrefauthors}%
Hallett, L\BPBI M.%
, Jones, S\BPBI K.%
, MacDonald, A\BPBI A\BPBI M.%
, Jones, M\BPBI B.%
, Flynn, D\BPBI F\BPBI B.%
, Ripplinger, J.%
\BDBL {}Collins, S\BPBI L.%
\end{APACrefauthors}%
\unskip\
\newblock
\APACrefYearMonthDay{2016}{}{}.
\newblock
{\BBOQ}\APACrefatitle {codyn: {An} r package of community dynamics metrics} {codyn: {An} r package of community dynamics metrics}.{\BBCQ}
\newblock
\APACjournalVolNumPages{Methods in Ecology and Evolution}{7}{10}{1146--1151}.
\PrintBackRefs{\CurrentBib}

\bibitem [\protect \citeauthoryear {%
Howard%
\ \BBA {} Hussain%
}{%
Howard%
\ \BBA {} Hussain%
}{%
{\protect \APACyear {2013}}%
}]{%
howard_democracys_2013}
\APACinsertmetastar {%
howard_democracys_2013}%
\begin{APACrefauthors}%
Howard, P\BPBI N.%
\BCBT {}\ \BBA {} Hussain, M\BPBI M.%
\end{APACrefauthors}%
\unskip\
\newblock
\APACrefYear{2013}.
\newblock
\APACrefbtitle {Democracy's {Fourth} {Wave}?: {Digital} {Media} and the {Arab} {Spring}} {Democracy's {Fourth} {Wave}?: {Digital} {Media} and the {Arab} {Spring}}.
\newblock
\APACaddressPublisher{}{Oxford University Press}.
\PrintBackRefs{\CurrentBib}

\bibitem [\protect \citeauthoryear {%
Kahan%
\ \protect \BOthers {.}}{%
Kahan%
\ \protect \BOthers {.}}{%
{\protect \APACyear {2012}}%
}]{%
kahan_polarizing_2012}
\APACinsertmetastar {%
kahan_polarizing_2012}%
\begin{APACrefauthors}%
Kahan, D\BPBI M.%
, Peters, E.%
, Wittlin, M.%
, Slovic, P.%
, Ouellette, L\BPBI L.%
, Braman, D.%
\BCBL {}\ \BBA {} Mandel, G.%
\end{APACrefauthors}%
\unskip\
\newblock
\APACrefYearMonthDay{2012}{{\APACmonth{10}}}{}.
\newblock
{\BBOQ}\APACrefatitle {The polarizing impact of science literacy and numeracy on perceived climate change risks} {The polarizing impact of science literacy and numeracy on perceived climate change risks}.{\BBCQ}
\newblock
\APACjournalVolNumPages{Nature Climate Change}{2}{10}{732--735}.
\PrintBackRefs{\CurrentBib}

\bibitem [\protect \citeauthoryear {%
Lin%
, Margolin%
, Keegan%
, Baronchelli%
\BCBL {}\ \BBA {} Lazer%
}{%
Lin%
\ \protect \BOthers {.}}{%
{\protect \APACyear {2013}}%
}]{%
lin_bigbirds_2013}
\APACinsertmetastar {%
lin_bigbirds_2013}%
\begin{APACrefauthors}%
Lin, Y\BHBI R.%
, Margolin, D.%
, Keegan, B.%
, Baronchelli, A.%
\BCBL {}\ \BBA {} Lazer, D.%
\end{APACrefauthors}%
\unskip\
\newblock
\APACrefYearMonthDay{2013}{}{}.
\newblock
{\BBOQ}\APACrefatitle {\# Bigbirds never die: Understanding social dynamics of emergent hashtags} {\# bigbirds never die: Understanding social dynamics of emergent hashtags}.{\BBCQ}
\newblock
\BIn{} \APACrefbtitle {Proceedings of the International AAAI Conference on Web and Social Media} {Proceedings of the international aaai conference on web and social media}\ (\BVOL~7, \BPGS\ 370--379).
\PrintBackRefs{\CurrentBib}

\bibitem [\protect \citeauthoryear {%
McElreath%
}{%
McElreath%
}{%
{\protect \APACyear {2016}}%
}]{%
mcelreath_statistical_2016}
\APACinsertmetastar {%
mcelreath_statistical_2016}%
\begin{APACrefauthors}%
McElreath, R.%
\end{APACrefauthors}%
\unskip\
\newblock
\APACrefYear{2016}.
\newblock
\APACrefbtitle {Statistical {Rethinking}: {A} {Bayesian} {Course} with {Examples} in {R} and {Stan}} {Statistical {Rethinking}: {A} {Bayesian} {Course} with {Examples} in {R} and {Stan}}.
\newblock
\APACaddressPublisher{New York}{Chapman and Hall/CRC}.
\newblock
\begin{APACrefDOI} \doi{10.1201/9781315372495} \end{APACrefDOI}
\PrintBackRefs{\CurrentBib}

\bibitem [\protect \citeauthoryear {%
Morrow%
, Bower%
\BCBL {}\ \BBA {} Greenspan%
}{%
Morrow%
\ \protect \BOthers {.}}{%
{\protect \APACyear {1989}}%
}]{%
morrow_updating_1989}
\APACinsertmetastar {%
morrow_updating_1989}%
\begin{APACrefauthors}%
Morrow, D\BPBI G.%
, Bower, G\BPBI H.%
\BCBL {}\ \BBA {} Greenspan, S\BPBI L.%
\end{APACrefauthors}%
\unskip\
\newblock
\APACrefYearMonthDay{1989}{}{}.
\newblock
{\BBOQ}\APACrefatitle {Updating situation models during narrative comprehension} {Updating situation models during narrative comprehension}.{\BBCQ}
\newblock
\APACjournalVolNumPages{Journal of memory and language}{28}{3}{292--312}.
\PrintBackRefs{\CurrentBib}

\bibitem [\protect \citeauthoryear {%
Papacharissi%
}{%
Papacharissi%
}{%
{\protect \APACyear {2015}}%
}]{%
papacharissi_affective_2015}
\APACinsertmetastar {%
papacharissi_affective_2015}%
\begin{APACrefauthors}%
Papacharissi, Z.%
\end{APACrefauthors}%
\unskip\
\newblock
\APACrefYear{2015}.
\newblock
\APACrefbtitle {Affective publics: sentiment, technology, and politics} {Affective publics: sentiment, technology, and politics}.
\newblock
\APACaddressPublisher{New York, NY}{Oxford University Press}.
\PrintBackRefs{\CurrentBib}

\bibitem [\protect \citeauthoryear {%
Papacharissi%
}{%
Papacharissi%
}{%
{\protect \APACyear {2016}}%
}]{%
papacharissi_affective_2016}
\APACinsertmetastar {%
papacharissi_affective_2016}%
\begin{APACrefauthors}%
Papacharissi, Z.%
\end{APACrefauthors}%
\unskip\
\newblock
\APACrefYearMonthDay{2016}{}{}.
\newblock
{\BBOQ}\APACrefatitle {Affective publics and structures of storytelling: sentiment, events and mediality} {Affective publics and structures of storytelling: sentiment, events and mediality}.{\BBCQ}
\newblock
\APACjournalVolNumPages{Information, Communication \& Society}{19}{3}{307--324}.
\PrintBackRefs{\CurrentBib}

\bibitem [\protect \citeauthoryear {%
Priniski%
, Verma%
\BCBL {}\ \BBA {} Morstatter%
}{%
Priniski%
\ \protect \BOthers {.}}{%
{\protect \APACyear {2023}}%
}]{%
priniski_pipeline_2023}
\APACinsertmetastar {%
priniski_pipeline_2023}%
\begin{APACrefauthors}%
Priniski, J.%
, Verma, I.%
\BCBL {}\ \BBA {} Morstatter, F.%
\end{APACrefauthors}%
\unskip\
\newblock
\APACrefYearMonthDay{2023}{}{}.
\newblock
{\BBOQ}\APACrefatitle {Pipeline for modeling causal beliefs from natural language} {Pipeline for modeling causal beliefs from natural language}.{\BBCQ}
\newblock
\BIn{} \APACrefbtitle {Proceedings of the 61st {Annual} {Meeting} of the {Association} for {Computational} {Linguistics} ({Volume} 3: {System} {Demonstrations})} {Proceedings of the 61st {Annual} {Meeting} of the {Association} for {Computational} {Linguistics} ({Volume} 3: {System} {Demonstrations})}\ (\BPGS\ 436--443).
\newblock
\APACaddressPublisher{Toronto, Canada}{Association for Computational Linguistics}.
\PrintBackRefs{\CurrentBib}

\bibitem [\protect \citeauthoryear {%
Sasahara%
\ \protect \BOthers {.}}{%
Sasahara%
\ \protect \BOthers {.}}{%
{\protect \APACyear {2021}}%
}]{%
sasahara_social_2021}
\APACinsertmetastar {%
sasahara_social_2021}%
\begin{APACrefauthors}%
Sasahara, K.%
, Chen, W.%
, Peng, H.%
, Ciampaglia, G\BPBI L.%
, Flammini, A.%
\BCBL {}\ \BBA {} Menczer, F.%
\end{APACrefauthors}%
\unskip\
\newblock
\APACrefYearMonthDay{2021}{}{}.
\newblock
{\BBOQ}\APACrefatitle {Social influence and unfollowing accelerate the emergence of echo chambers} {Social influence and unfollowing accelerate the emergence of echo chambers}.{\BBCQ}
\newblock
\APACjournalVolNumPages{Journal of Computational Social Science}{4}{1}{381--402}.
\PrintBackRefs{\CurrentBib}

\bibitem [\protect \citeauthoryear {%
Tangherlini%
, Shahsavari%
, Shahbazi%
, Ebrahimzadeh%
\BCBL {}\ \BBA {} Roychowdhury%
}{%
Tangherlini%
\ \protect \BOthers {.}}{%
{\protect \APACyear {2020}}%
}]{%
tangherlini_automated_2020}
\APACinsertmetastar {%
tangherlini_automated_2020}%
\begin{APACrefauthors}%
Tangherlini, T\BPBI R.%
, Shahsavari, S.%
, Shahbazi, B.%
, Ebrahimzadeh, E.%
\BCBL {}\ \BBA {} Roychowdhury, V.%
\end{APACrefauthors}%
\unskip\
\newblock
\APACrefYearMonthDay{2020}{{\APACmonth{06}}}{}.
\newblock
{\BBOQ}\APACrefatitle {An automated pipeline for the discovery of conspiracy and conspiracy theory narrative frameworks: {Bridgegate}, {Pizzagate} and storytelling on the web} {An automated pipeline for the discovery of conspiracy and conspiracy theory narrative frameworks: {Bridgegate}, {Pizzagate} and storytelling on the web}.{\BBCQ}
\newblock
\APACjournalVolNumPages{PLOS ONE}{15}{6}{e0233879}.
\PrintBackRefs{\CurrentBib}

\bibitem [\protect \citeauthoryear {%
Yang%
}{%
Yang%
}{%
{\protect \APACyear {2016}}%
}]{%
yang_narrative_2016}
\APACinsertmetastar {%
yang_narrative_2016}%
\begin{APACrefauthors}%
Yang, G.%
\end{APACrefauthors}%
\unskip\
\newblock
\APACrefYearMonthDay{2016}{}{}.
\newblock
{\BBOQ}\APACrefatitle {Narrative {Agency} in {Hashtag} {Activism}: {The} {Case} of \#{BlackLivesMatter}} {Narrative {Agency} in {Hashtag} {Activism}: {The} {Case} of \#{BlackLivesMatter}}.{\BBCQ}
\newblock
\APACjournalVolNumPages{Media and Communication}{4}{4}{13--17}.
\PrintBackRefs{\CurrentBib}

\bibitem [\protect \citeauthoryear {%
Zwaan%
, Langston%
\BCBL {}\ \BBA {} Graesser%
}{%
Zwaan%
, Langston%
\BCBL {}\ \BBA {} Graesser%
}{%
{\protect \APACyear {1995}}%
}]{%
zwaan_construction_1995}
\APACinsertmetastar {%
zwaan_construction_1995}%
\begin{APACrefauthors}%
Zwaan, R\BPBI A.%
, Langston, M\BPBI C.%
\BCBL {}\ \BBA {} Graesser, A\BPBI C.%
\end{APACrefauthors}%
\unskip\
\newblock
\APACrefYearMonthDay{1995}{}{}.
\newblock
{\BBOQ}\APACrefatitle {The construction of situation models in narrative comprehension: An event-indexing model} {The construction of situation models in narrative comprehension: An event-indexing model}.{\BBCQ}
\newblock
\APACjournalVolNumPages{Psychological science}{6}{5}{292--297}.
\PrintBackRefs{\CurrentBib}

\bibitem [\protect \citeauthoryear {%
Zwaan%
, Magliano%
\BCBL {}\ \BBA {} Graesser%
}{%
Zwaan%
, Magliano%
\BCBL {}\ \BBA {} Graesser%
}{%
{\protect \APACyear {1995}}%
}]{%
zwaan_dimensions_1995}
\APACinsertmetastar {%
zwaan_dimensions_1995}%
\begin{APACrefauthors}%
Zwaan, R\BPBI A.%
, Magliano, J\BPBI P.%
\BCBL {}\ \BBA {} Graesser, A\BPBI C.%
\end{APACrefauthors}%
\unskip\
\newblock
\APACrefYearMonthDay{1995}{}{}.
\newblock
{\BBOQ}\APACrefatitle {Dimensions of situation model construction in narrative comprehension.} {Dimensions of situation model construction in narrative comprehension.}{\BBCQ}
\newblock
\APACjournalVolNumPages{Journal of experimental psychology: Learning, memory, and cognition}{21}{2}{386}.
\PrintBackRefs{\CurrentBib}

\bibitem [\protect \citeauthoryear {%
Zwaan%
\ \BBA {} Radvansky%
}{%
Zwaan%
\ \BBA {} Radvansky%
}{%
{\protect \APACyear {1998}}%
}]{%
zwaan_situation_1998}
\APACinsertmetastar {%
zwaan_situation_1998}%
\begin{APACrefauthors}%
Zwaan, R\BPBI A.%
\BCBT {}\ \BBA {} Radvansky, G\BPBI A.%
\end{APACrefauthors}%
\unskip\
\newblock
\APACrefYearMonthDay{1998}{}{}.
\newblock
{\BBOQ}\APACrefatitle {Situation models in language comprehension and memory.} {Situation models in language comprehension and memory.}{\BBCQ}
\newblock
\APACjournalVolNumPages{Psychological bulletin}{123}{2}{162}.
\PrintBackRefs{\CurrentBib}

\end{thebibliography}

\end{document}